\documentstyle[aps,prl,multicol,epsf]{revtex}
\global\firstfigfalse
\global\firsttabfalse

\newcommand{\bleq}{\ifpreprintsty
                   \else
                   \end{multicols}\vspace*{-3.5ex}{\tiny
                   \noindent\begin{tabular}[t]{c|}
                   \parbox{0.493\hsize}{~} \\ \hline \end{tabular}}
                   \fi}
\newcommand{\eleq}{\ifpreprintsty
                   \else
                   {\tiny\hspace*{\fill}\begin{tabular}[t]{|c}\hline
                    \parbox{0.49\hsize}{~} \\
                    \end{tabular}}\vspace*{-2.5ex}\begin{multicols}{2}
                    \fi}
\newcommand{\bcols}{\ifpreprintsty\else\begin{multicols}{2}\fi}
\newcommand{\ecols}{\ifpreprintsty\else\end{multicols}\fi}

\begin{document}
\bibliographystyle{prsty}
\title{ Quantum Pumping in the Magnetic Field: Role of Discrete Symmetries}
  
\draft

\author{I.L. Aleiner$^{1}$, B.~L.~Altshuler$^{2}$, and A.~Kamenev$^{3}$}
\address{$^{1}$ Department of Physics and Astronomy, SUNY at Stony
Brook, Stony Brook, NY 11794\\ $^{2}$ Physics Department, Princeton
University, NJ 08544\\ and NEC Research Institute, 4 Independence Way,
Princeton, NJ 08540\\ $^{3}$ Department of Physics, Technion, Haifa
32000, Israel.  \\ {}~{\rm (\today)}~ \medskip \\ 
\parbox{14cm} {\rm We consider an effect of the discrete spatial
symmetries and magnetic field on the adiabatic charge pumping in
mesoscopic systems. In general case, there is no symmetry of the
pumped charge with respect to the inversion of magnetic field $Q(B)
\neq Q(-B)$. We find that the reflection symmetries  give
rise to relations $Q(B)=Q(-B)$ or $Q(B)=-Q(-B)$ depending 
on the orientation 
of the reflection axis. In  presence
of the center of inversion, $Q(B)\equiv 0$. Additional symmetries 
may arise in the case of bilinear pumping. 
\smallskip\\ 
PACS numbers: 72.10.Bg, 73.22.-b, 05.45.+b }\bigskip \\ }

\maketitle

\bcols
The phenomenon of
adiabatic charge pumping has attracted considerable theoretical and
experimental interest during the last decade
\cite{Thouless83,Kouwenhoven,Aleiner98,Brouwer98,Zhou99,Shutenko,Marcus,Andreev00,avron}.
It occurs when the Hamiltonian of the system is changed
periodically with time: under certain condition a finite charge my be
transmitted through the system during each period of the oscillation.
Such a charge transfer
takes place even if no dc voltage is applied.
 The idea is originally due to Thouless \cite{Thouless83}, who
showed that in some one-dimensional systems the transmitted charge
is quantized in the adiabatic limit.

Such quantization is possible only if the two-terminal conductance of
the system vanishes. It was shown \cite{Aleiner98,Shutenko,Andreev00}
that in addition to the quantized part of the charge transfer
\cite{Thouless83} there is another contribution which is proportional
to the dissipative conductance of the system.  As a result, in
mesoscopic system, the charge (i) is not quantized and (ii) exhibits
strong sample to sample fluctuations \cite{foot1}.

Since the adiabatic pumping is a phase coherent mesoscopic effect it
may be strongly sensitive to an external magnetic
field, $B$. The qualitatively important aspect is the presence (or
absence) of any symmetry relations upon reversing the sign of $B$.  It
was suggested theoretically \cite{Zhou99} that the transmitted charge,
$Q$, is invariant upon such field reversal
\begin{equation} 
                                \label{symm}
Q(B) = Q(-B)\, .
\end{equation}
This relation, if true, would be an analog of the famous Onsager
symmetry for the two--terminal conductance: $G(B)= G(-B)$.  The subsequent
experiment \cite{Marcus} on mesoscopic quantum dots appeared to be in
a good agreement with Eq.~(\ref{symm}).  It soon became clear,
however, \cite{Shutenko} that there is no real theoretical
justification for the symmetry relation like Eq.~(\ref{symm}). Indeed,
unlike the conductance which is determined only by the  moduli of the
transmission eigenvalues, the pumped charge
involves eigenfunctions as well.  The latter, in general, do not obey
any symmetry relation and so does not transmitted charge.  As a
result, the experimental confirmation of Eq.~(\ref{symm})
in Ref.~\cite{Marcus} appears to be a puzzle.

\begin{figure}
\vglue 0cm
\hspace{0.01\hsize}
\epsfxsize=0.8\hsize
\epsffile{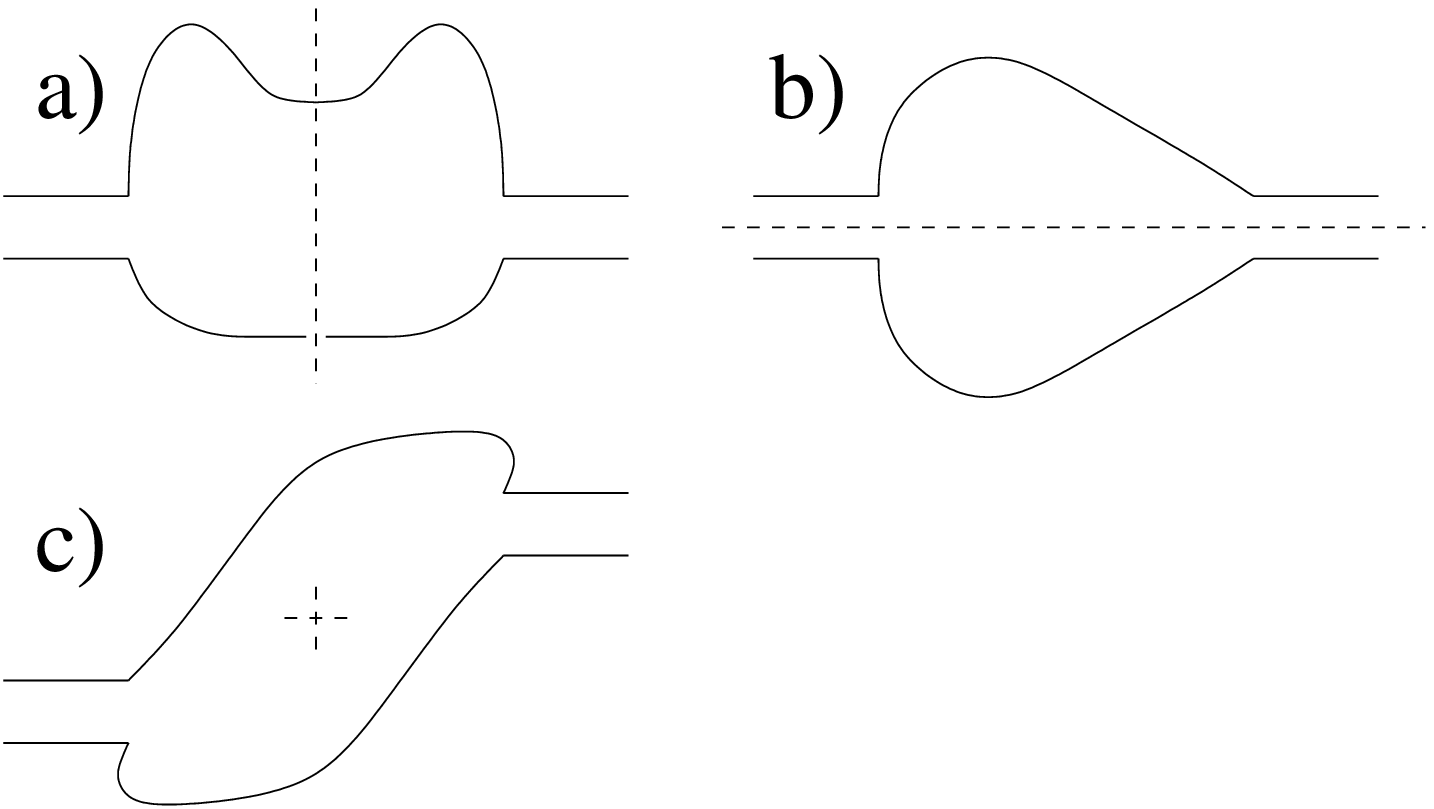}
\refstepcounter{figure} \label{fig1}
{\\ \small FIG.\ \ref{fig1}
Shematic representation of the  possible reflection symmetries: 
(a) left--right (LR); (b) up--down (UD);
(c) inversion (I).  \par}
\end{figure}

The purpose of this work is two--fold.
First, we intend to demonstrate explicitly the role of the
dissipation in lifting the symmetry with respect to the magnetic
field inversion. 
Second, we 
 consider how discrete spatial
symmetries which the quantum dot may have manifest themselves in 
 the magnetic symmetries of the
transmitted charge.  

It has been known for some time that the
 discrete symmetries may change level statistics
\cite{Robnik86} as well as influence quantum correction to the
transport coefficients \cite{Baranger97}. It is therefore natural to
explore their effect on the adiabatic pumping and in particular its
magneto--dependence.  A quantum dot with two leads may
posses three distinct types of spatial symmetries 
which, following Ref.~\cite{Baranger97}, we call LR, UD, and I, 
see Fig.~\ref{fig1}.  If these spatial
symmetries are kept intact in the pumping cycle, they give rise to
definite magnetic symmetries of the transmitted charge. Curiously the
corresponding magnetic symmetry are qualitatively distinct. For the UD
symmetry we find 
\begin{equation} 
                                \label{UD}
Q_{UD}(B) = Q_{UD}(-B)\, ,
\end{equation}
in agreement with an earlier speculation 
\cite{Zhou99}
and experimental results
\cite{Marcus}.
On the other hand,  dots with LR symmetry obey  a qualitatively 
different  relation  
\begin{equation} 
                                \label{LR}
Q_{LR}(B) = - Q_{LR}(-B)\, .
\end{equation}
When both of the symmetries are present, one finds
that the only possibility consistent  with both Eqs.~(\ref{UD}) and (\ref{LR})
is \begin{equation}
Q(B) = 0, 
\label{I}
\end{equation}
  The same
relation holds under the weaker condition that 
 the dot  has the center of inversion (I),
see Fig.~\ref{fig1}(c).
Finally, dots without any spatial symmetry, in general, do not have
any definite relation between $Q(B)$ and $Q(-B)$.

Let us  present a formal  proof of the advertised results.
We consider  a phase coherent scattering region  
connected to the two leads, each having $n$ transverse channels. 
Such system may be described by 
a $ 2n\times  2n$ unitary scattering matrix, $\hat S$, 
of the following structure 
\begin{equation} 
\hat S =\pmatrix{  
\hat{r} & \hat{t}' \cr
\hat{t} & \hat{r}' 
}
\, ,
                                                             \label{Smat}
\end{equation}
where $\hat{r}$ ($\hat{r}'$) and $\hat{t}$ ($\hat{t}'$) are $n\times
n$ left (right) reflection and transmission matrices
correspondingly. This matrix can be diagonalized simultaneously by the
{\em block-diagonal} unitary matrices $\hat U$ and $\hat V$ (see
e.g. Ref.~\cite{Stone95})
\begin{equation} 
\hat S = \hat U  {\tilde S}  \hat V^{\dagger}\, .
                                                             \label{tdep}
\end{equation}
Here ${\tilde S}$ is a matrix of the form Eq.~(\ref{Smat}) with real
diagonal reflection and transmission blocks: $\tilde t = \tilde t'=
\mbox{diag}\{t_1,\ldots t_n\}$ and $\tilde r = -\tilde r'=
\mbox{diag}\{\sqrt{1-t_1^2},\ldots \sqrt{1-t_n^2}\}$, where $t_i^2$
are the transmission coefficients. Matrices $\hat U$ and $\hat V$
 satisfy the relation
\begin{equation} 
\left[\hat U, \hat \sigma_z \right]
= \left[\hat V, \hat \sigma_z \right]
=0.                                                             \label{commute}
\end{equation}
The Pauli matrices $\hat{\sigma}_i$ are defined as 
\begin{equation} 
\hat{\sigma}_0=\pmatrix{  
\hat{I} & 0 \cr
0 & \hat{I} }
\, , \,\,\,\, 
\hat{\sigma}_z=\pmatrix{  
\hat{I} & 0 \cr
0 & - \hat{I}  } 
\, , \,\,\,\,      
\hat{\sigma}_x=\pmatrix{  
0 & \hat{I} \cr
\hat{I} & 0 }
\, ,
                                                       \label{Pauli}
\end{equation}
where the $2\times 2$ structure represents the  space of left--right leads 
(the same as in Eq.~(\ref{Smat}) ) and  
$\hat{I}$ is the unit $n\times n$ matrix.

In these notations, the Landauer formula for the two--terminal
conductance can be written as
\begin{equation} 
G ={e^2\over 2\pi \hbar}\, 
\mbox{Tr}\{ \hat S^{\dagger}\hat \sigma_l \hat S\hat \sigma_r\}=
{e^2\over 8\pi \hbar}\, 
\mbox{Tr}\left\{
\hat \sigma_z
\left[\hat \sigma_z - 
 \tilde S^{\dagger}\hat \sigma_z \tilde S
\right]
\right\}\, ,
                                                             \label{land}
\end{equation}
where $\hat \sigma_{l/r}\equiv (\hat \sigma_0 \pm\hat \sigma_z)/2 $
are projectors onto the left (right) leads. The last equality in
Eq.~(\ref{land}) utilizes Eq.~(\ref{commute}) and hence the phase
matrices $\hat U$ and $\hat V$ drop out from the
transport coefficients.

If the $\hat S$--matrix is a periodic function of time $\tau$ with the period
$\tau_0$, $\hat S(\tau+\tau_0) =\hat S(\tau)$, a certain amount of charge may be
transferred through the scattering region upon the completion of each
cycle.  Provided that the time dependence is adiabatic, i.e. $\hat S(\tau)$
is slow on the scale of the
Wigner delay time, the average transmitted charge is given by
\cite{Brouwer98}
\begin{equation} 
 Q  =
{e\over 2i} \int\limits_0^{\tau_0}\!\! {d \tau\over 2\pi}\,   
\mbox{Tr}    \left\{
{\partial \hat S\over\partial \tau} \hat S^{\dagger} \hat \sigma_z 
\right\}\, .
                                                             \label{aver}
\end{equation}
It may be rewritten as a sum of the two contributions $Q=Q_1 + Q_2$, where
an anomalous 
\cite{Aleiner98,Shutenko,Andreev00}
contribution $Q_1$ results in  a quantized charge   
\begin{equation} 
Q_1 = {e\over 2i} \int\limits_0^{\tau_0}\!\! {d \tau\over 2\pi}\,  
\mbox{Tr} \left\{ 
 \hat U^{\dagger} {\partial \hat U\over\partial \tau}\hat \sigma_z  - 
 \hat V^{\dagger} {\partial \hat V\over\partial \tau}\hat \sigma_z  
\right\} \,  . 
                                                             \label{Q1}
\end{equation}
The second contribution, $Q_2$, is {\em not} quantized, and given by
\begin{equation} 
Q_2 = {e\over 2i} \int\limits_0^{\tau_0}\!\! {d \tau\over 2\pi}\,  
\mbox{Tr} \left\{
 \hat V^{\dagger} {\partial \hat V\over\partial \tau}
\left[ 
\hat \sigma_z -  
\tilde S^{\dagger} \hat \sigma_z  \tilde S
\right] 
\right\} \,  . 
                                                             \label{Q2}
\end{equation}
It is important to emphasize that the factor in brackets coincides with
that in Eq.~(\ref{land}). This means that the contribution $Q_2$ is
determined by the dissipation in the leads in the same way as the Landauer
conductance is.
Contribution $Q_2$ vanishes for non--transparent dots, $\tilde t=0$, or in the
presence of gap in the excitation spectrum. 
Note also that the time
derivative of $\tilde S$ does not contribute to the average
transmitted charge.

The 
$\hat S$--matrix must be invariant upon simultaneous transposition and 
magnetic field inversion \cite{Landau}:  
\begin{equation} 
\hat S(-B) = \hat S^T(B)\, .
                                                             \label{BB}
\end{equation}
In terms of the decomposition, Eq.~(\ref{tdep}), this symmetry implies
that
\begin{equation} 
\hat U(-B) = \hat V^*(B),\quad \hat V(-B)= \hat U^*(B), \quad \tilde
S(-B) = \tilde S(B).
\end{equation}
 The last equality combined with Eq.~(\ref{land}) immediately yields
the Onsager relation for the conductance, $G(-B) = G(B)$. Applying
these relation to the adiabatic pumping, one finds that the quantized
component of the transmitted charge, $Q_1$, is a symmetric function of
the magnetic field, $Q_1(-B) = Q_1(B)$. On the other hand, the
dissipative component of the pumped charge, $Q_2$, in general, does
{\em not} exhibit any definite symmetry upon field reversal. Therefore,
the experimental finding \cite{Marcus} of the magnetic field symmetric
pumping in open quantum dot requires some additional understanding. In
what follows we show that spatial symmetries of the dot indeed can enforce
 magnetic symmetries.

Consider a two--fold spatial symmetry operation,  $\hat O$, such that 
$\hat O^2$ =1. Upon this symmetry the $\hat S$--matrix transforms according to
\begin{equation} 
\hat S \to \hat O\hat S\hat O^{-1} = \hat O\hat S\hat O\, .
                                                             \label{osymm}
\end{equation}   
The system conductance, as well as shot--noise and higher order
current correlators, must remain invariant under such
transformation. This means that
$\mbox{Tr}\{ (\hat S^{\dagger}\hat \sigma_l \hat S\hat \sigma_r)^k\}$,
 is invariant under transformation (\ref{osymm}),
for any positive integer $k$.  
Such  invariance is possible only if one of the two conditions is fulfilled:
either 
\begin{mathletters}
\begin{equation}
\hat O\hat \sigma_z  = \hat \sigma_z\hat O
\label{sym1} 
\end{equation}
(consequently $\hat \sigma_{l/r} \to
\hat \sigma_{l/r}$ meaning that each lead is transformed to itself upon the
symmetry transformation), or 
\begin{equation}
\hat O\hat \sigma_z  = - \hat \sigma_z\hat O
\label{sym2} 
\end{equation}
(the leads are
interchanged by the symmetry, $\hat \sigma_{l/r} \to \hat \sigma_{r/l}$).
\end{mathletters}

An example of the  symmetry (\ref{sym1}) is reflection relative to the axis 
connecting the two leads, see Fig. \ref{fig1}(b), i.e.  
UD symmetry. All the channels may be classified by the parity of their 
wavefunctions with respect to inversion around the reflection axis.
The corresponding $\hat S$--matrix at $B=0$ acquires a block--diagonal structure in 
the space of even--odd channels. 
In other words,
\[
\hat{S}(B=0)= \hat \Sigma^{eo}_z\hat{S}(B=0) \hat \Sigma^{eo}_z,
\]
where $\hat \Sigma^{eo}_z$ is a diagonal $2n \times 2n$ matrix
$\left[\hat \Sigma^{eo}_z\right]_{ik}=  p_i\delta_{ik}$, and
$p_i=1\, (-1)$ for the 
even (odd) $i$th channel.  If the magnetic field $B$ is applied to the
system, one can write the Hamiltonian in the Landau gauge as $(p_x -
By)^2 + p_y^2 + V_{UD}(x,y)$, where the $x$--axis is the symmetry
axis, $V_{UD}(x,y)=V_{UD}(x,-y)$.  It is easy to see that the
Hamiltonian is invariant under simultaneous inversion of the magnetic
field and reflection with respect to the $x$--axis. As a result
\begin{equation} 
\hat{S}(-B) = \hat\Sigma^{eo}_z \hat{S}(B) \hat\Sigma^{eo}_z \, ,
                                                             \label{UDS}
\end{equation}  
and therefore $\hat V (B) = \hat \Sigma^{eo}_z \hat V (-B)$,
$\hat U (B) = \hat U (-B)\hat \Sigma^{eo}_z$.
Substituting  these relations into Eq.~(\ref{Q2}), 
one finds that the dissipative component of pumped charge is magnetic field 
symmetric, $Q_2(B) = Q_2(-B)$. Since the quantized component $Q_1$ is always 
symmetric, one concludes that for the UD reflection symmetry the 
transmitted  charge obeys  Eq.~(\ref{UD}).

An example of the symmetry of type (\ref{sym2}) is the reflection
symmetry depicted on Fig.~\ref{fig1}(a). We call it LR symmetry.  In
this case $\hat O=\hat\sigma_x$, where $\hat \sigma_x$ is defined in
Eq.~(\ref{Pauli}).  In the appropriate gauge  the Hamiltonian 
takes the form 
$p_x^2 + (p_y + Bx)^2 + V_{LR}(x,y)$ with $V_{LR}(x,y)= V_{LR}(-x,y)$. 
The system is invariant under 
inversion of the magnetic field direction and reflection with respect 
to the $y$--axis simultaneously. As a result one obtains
\begin{equation} 
\hat S(-B) = \hat \sigma_x \hat S(B) \hat \sigma_x \, .
                                                             \label{LRS}
\end{equation} 
Employing Eq.~(\ref{aver}) and the fact that $\hat \sigma_x \hat \sigma_z
\hat \sigma_x = -\hat \sigma_z$, one finds $Q(-B) = - Q(B)$, as was announced in
Eq.~(\ref{LR}). Since the quantized component is always symmetric, it
must be absent for the LR symmetry. The remaining pure dissipative
component, $Q_2$, is antisymmetric in this case. In the absence of the
field, $Q(B=0) = 0$, which is obviously true in case of the LR
symmetry, because the two directions of the current flow are
equivalent.

If the dot exhibits both UD and LR symmetries (four--fold symmetry in
the terminology of Ref.~ \cite{Baranger97}) which are preserved in the
pumping process, then the pumped charge vanishes, $Q(B)\equiv
0$. Indeed this is the only possibility to satisfy both
Eqs.~(\ref{UD}) and (\ref{LR}). The same is true if the dot has a
center of inversion, Fig.~\ref{fig1}(c).  
In this case the Hamiltonian is invariant upon
changing direction of both $x$ and $y$ axis, but keeping $B$
intact. As a result,
\[
\hat \sigma_x
\hat \Sigma^{eo}_z  \hat S(B)\hat \Sigma^{eo}_z \hat \sigma_x = \hat S(B).
\]  
Employing Eq.~(\ref{aver}), one obtains $Q=-Q=0$.

Due to the presence of the dissipative contribution $Q_2$, see
Eq.~(\ref{Q2}), the transmitted charge fluctuates from cycle to cycle;
the resulting random process can be characterized by the distribution
function $P(Q)$. 
Using the formalism of Ref.~\cite{Andreev00}, one may show
that not only the average charge, but the entire distribution
function, $P(Q)$, exhibits magnetic symmetries if the dot is spatially
symmetric. Namely, for the case of the UD symmetry one easily obtains
$P(Q,B) = P(Q,-B)$, for the LR symmetry $P(Q,B) = P(-Q,-B)$, and
for the inversion symmetry $P(Q,B) = P(-Q,B)$ and for four-fold
symmetry $P(Q,B) = P(-Q,B)= P(Q,-B) $.

So far we considered the pumping of arbitrary strength however
preserving the initial symmetry of the dot. In the case of so-called
bilinear response \cite{Brouwer98,Zhou99} 
one may make some conclusions even for perturbations
violating the symmetry.
Consider the time dependent potential of the form
\begin{equation}
V(x,y;t)=V(x,y) + X_1(t)V_1(x,y)+X_2(t)V_1(x,y)
\label{pot}
\end{equation}
the last two terms describe  potential profiles  created by the pumps.
In the lowest non-vanishing order in $X_{1,2}$, Eq.~(\ref{aver}) takes
the form
\begin{equation}
Q^{bl}  =
{e\over 2\pi} {\cal A}_{X}\, {\mathrm Im} 
\left.
 \mbox{Tr} \left\{
{\partial \hat S\over\partial X_1} 
{\partial \hat S^\dagger\over\partial X_2}
 \hat \sigma_z 
\right\}
\right|_{X_{1,2}=0}
,
\label{bl}
\end{equation}
where ${\cal A}_{X}$ is the area on the $(X_1,X_2)$ plane enclosed by
the contour $[X_1(t),X_2(t)]$.

Let the unperturbed system be UD-symmetric, i.e.
$V(x,y)=V(x,-y)$. We wish to consider the perturbations within
irreducible representation of the discrete symmetry group, i.e.
either symmetric or antisymmetric: $V_i(x,y)=p_i V_i(x,-y)$, 
where $p_i=\pm 1$.
We have already considered the case of $p_1=p_2=1$, i.e. not changing
the symmetry of the dot, see Eq.~(\ref{UD}). The question we are going to
address now is what happens if at least one of the factors $p_i$
is negative.
Similar problem can be posed for the $LR$ symmetric dot:
$V(x,y)=V(-x,y)$ and $V_i(x,y)=p_i V_i(-x,y)$ 

Analogously to Eqs.~(\ref{UDS}) and (\ref{LRS}), one obtains
\begin{eqnarray*}
{\partial\hat S(-B) \over X_i} 
= p_i
\hat \sigma_x {\partial\hat S(B) \over X_i}\hat \sigma_x, \quad (LR);\\
{\partial\hat S(-B) \over X_i} = p_i\hat\Sigma^{eo}_z  {\partial\hat S(B) \over X_i}
 \hat\Sigma^{eo}_z,
\quad (UD).
\end{eqnarray*} 
Thus, we find instead of Eqs.~(\ref{UD}) and (\ref{LR})
\begin{mathletters}
\label{blsd}
\begin{eqnarray}
Q_{UD}^{bl}(B) = p_1p_2Q_{UD}^{bl}(-B),\\
Q_{LR}^{bl}(B) = -p_1p_2Q_{LR}^{bl}(-B).
\end{eqnarray}
\end{mathletters}
To avoid a confusion, notice  that Eqs.~(\ref{blsd}) are not valid
beyond the bilinear response approximation if at least one of $p_i=-1$.

To conclude, we have demonstrated that a general adiabatic pump does
not obey any symmetry with respect to magnetic field reversal.  Such
symmetry, however, may be  recovered as a result of the discrete spatial
symmetry of the pump. Moreover, different spatial symmetries lead to a
qualitatively distinct behavior in the magnetic field.  The average
transmitted charge (and its higher moments) is a symmetric or
antisymmetric (or identically zero) function of the field depending on
the type of spatial symmetry of the dot. Our findings may possibly
clarify the origin of the experimentally observed magnetic symmetry
\cite{Marcus}, as well as motivate further experiments.

We appreciate the warm hospitality of the Norwegian Center for
Advanced Studies, where part of this work was performed.  
A.K. was partly supported by  BSF-9800338 grant. 
I.~A. is a Packard Fellow.

\ecols
\end{document}